\documentclass[12pt]{emulateapj}
\newcommand{\teff}{$T_{\rm eff}$}
\newcommand{\tiff}{T_{\rm eff}}

\newcommand{\rwd}{J1102+4113}
\journalinfo{Submitted to AJ Feb. 22, 2008; Accepted Apr. 9, 2008}
\slugcomment{}
\shorttitle{Old Halo WD Candidate}
\shortauthors{Hall et al.}

\begin{document}
\title{A Nearby Old Halo White Dwarf Candidate from the Sloan Digital Sky Survey}
\author{
Patrick B. Hall,\altaffilmark{1} 
Piotr M. Kowalski,\altaffilmark{2}
Hugh C. Harris,\altaffilmark{3} 
Akshay Awal,\altaffilmark{1,4} 
S. K. Leggett,\altaffilmark{5}
Mukremin Kilic,\altaffilmark{6}
Scott F. Anderson,\altaffilmark{7}
Evalyn Gates\altaffilmark{8}
}
\altaffiltext{1}{Department of Physics and Astronomy,
York University, Toronto, Ontario M3J 1P3, Canada}
\altaffiltext{2}{Lehrstuhl f\"ur Theoretische Chemie, Ruhr-Universit\"at, 44780 Bochum, Germany}
\altaffiltext{3}{United States Naval Observatory, Flagstaff, AZ 86001, USA}
\altaffiltext{4}{Emery Collegiate Institute, Toronto, Ontario M9M 2V9, Canada}
\altaffiltext{5}{Gemini Observatory, 670 North A'ohoku Place Hilo, HI 96720, USA}
\altaffiltext{6}{Department of Astronomy, The Ohio State University, Columbus, OH 43210, USA}
\altaffiltext{7}{Department of Astronomy, University of Washington, Seattle, WA 98195, USA}
\altaffiltext{8}{Kavli Institute for Cosmological Physics, The University of Chicago, Chicago, IL 60637, USA}

\begin{abstract}
We report the discovery of a nearby, old, halo white dwarf candidate from the
Sloan Digital Sky Survey.  SDSS J110217.48+411315.4 has a proper motion of
1.75$\arcsec$/year and redder optical colors than all other known featureless
(type DC) white dwarfs.  We present SDSS imaging and spectroscopy of this
object, along with 
near-infrared photometry obtained at the United Kingdom Infra-Red Telescope. 
Fitting its photometry with up-to-date model atmospheres, we find that its
overall spectral energy distribution is fit reasonably well with a pure hydrogen
composition and $\tiff\approx 3800$~K (assuming $\log g=8$).  That temperature
and gravity would place this white dwarf at 35 pc from the Sun with a tangential
velocity of 290 km s$^{-1}$ and space velocities consistent with halo 
membership; furthermore, its combined main sequence and white dwarf cooling age
would be $\approx$11 Gyr.  However, if this object is a massive white dwarf, 
it could be a younger object with a thick disk origin.
Whatever its origin, the optical colors of this object are redder than 
predicted by any current pure hydrogen, pure helium or mixed hydrogen-helium
atmospheric model, indicating that there remain problems in our understanding 
of the complicated physics of the dense atmospheres of cool white dwarfs.
\end{abstract}

\section{Introduction} \label{intro}

There is considerable interest in white dwarf (WD) stellar remnants
in the halo of our Galaxy.
Since halo stars are generally older than disk stars, the oldest halo WDs 
should be older and cooler than the oldest disk WDs.  
Such objects are of interest for studying the age distribution of halo stars
and for testing our understanding of stellar atmospheres, since their spectra
can depart significantly from simple blackbodies 
(see, e.g., {Hansen} 1998; {Saumon} \& {Jacobson} 1999; {Kowalski} 2006a).
Halo WD candidates may be identifiable as faint objects with high proper
motions and unusual colors (e.g., {Ducourant} {et~al.} 2007).
Some halo WD candidates have indeed been found, but most of them have 
relatively warm temperatures, indicating they are relatively young white dwarfs 
(e.g., {Bergeron} {et~al.} 2005; {L{\'e}pine}, {Rich}, \& {Shara} 2005).  To date, the coolest known {\em probable}
halo WD is WD~0346+246, with $T\simeq 3800$~K ({Bergeron} 2001).  
The coolest halo WD {\em candidate}, whose velocities are consistent either 
with a halo or a thick disk origin, is SDSS J122048.65+091412.1 ({Gates} {et~al.} 2004). 
The latter object is one of a handful of {\em ultracool} WDs
($\tiff\lesssim 3800$~K) whose optical spectra show collision-induced 
absorption (CIA) from H$_2$.  
CIA causes WDs to exhibit increasingly bluer colors at increasingly
shorter wavelengths at $\tiff\lesssim 5000$~K (see, e.g., {Bergeron} {et~al.} 2005).

Here we report the identification of a candidate old, halo WD with red optical
colors whose spectrum was obtained by the Sloan Digital Sky Survey 
(SDSS; {York} {et~al.} 2000).  The SDSS used a drift-scanning
imaging camera ({Gunn} {et~al.} 1998) on a 2.5-m telescope ({Gunn} {et~al.} 2006) to image
$\sim$10$^4$\,deg$^2$ of sky on the SDSS $ugriz$
magnitude system ({Fukugita} {et~al.} 1996; {Hogg} {et~al.} 2001; {Smith} {et~al.} 2002; {Pier} {et~al.} 2003; {Ivezi{\'c}} {et~al.} 2004; {Tucker} {et~al.} 2006).
Two multi-fiber, double spectrographs are being used to obtain $R\sim2100$
spectra for $\sim$10$^6$ galaxies
and $\sim$10$^5$ quasar candidates ({Stoughton} {et~al.} 2002).
As discussed in {Richards} {et~al.} (2002), most quasar candidates are targeted for
spectroscopy because they are outliers from the stellar locus.
Spectroscopy of such targets provides data not just on quasars, but also on
objects with colors different from those of the stellar locus, such as 
the unusual white dwarf presented herein.

\section{Images and Photometry} \label{imgs}

\begin{figure} \epsscale{0.95} \plotone{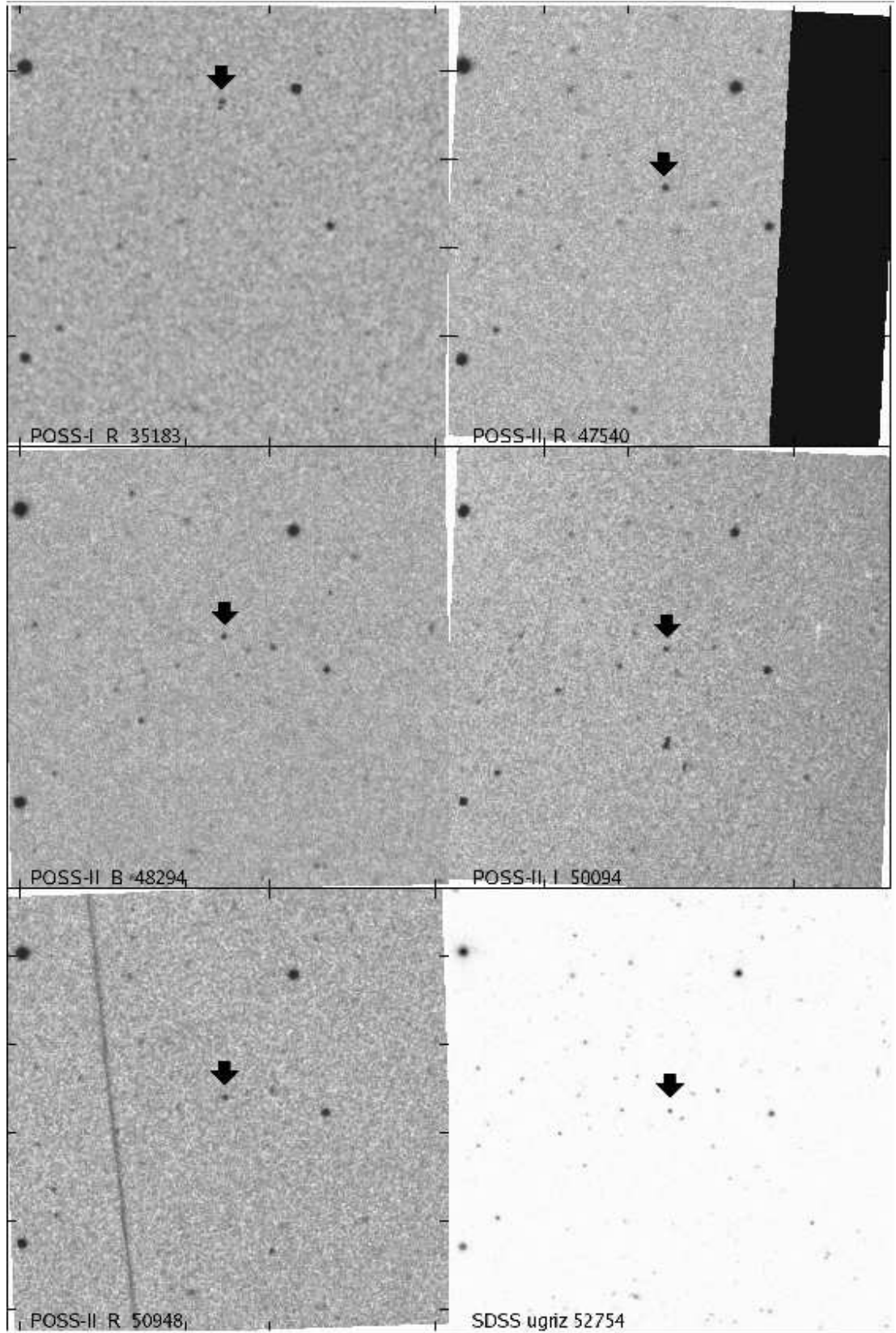} 
\caption{Imaging of \rwd\ (just below each arrow) at six different epochs,
centered on its position in the epoch of the SDSS image at lower right.  The 
image source, equivalent passband, and Modified Julian Date (MJD) are indicated
in each image.
\rwd\ moves from the top middle to the center of the images over time.
Each image is 5\arcmin$\times$5\arcmin\ in size, with North up and East at left.
}\label{f_imgs} \end{figure}

SDSS J110217.48+411315.4 (hereafter J1102+4113)\footnote{Due 
to its high proper motion, the SDSS coordinates for this object are
in Equinox J2000 and Epoch 2003.3.  The United States Naval Observatory 
USNO-B1.0 coordinates for this object
in Equinox J2000 and Epoch 2000 are RA=11:02:17.50 and DEC=+41:13:21.52.}
was noticed during visual
inspection of all SDSS spectra from Data Release Six ({Adelman-McCarthy} {et~al.} 2008)
classified as {\sc unknown} by the SDSS pipeline.
It stood out in the SDSS Catalog Archive Server
as having a high proper motion 
of $\mu_{\alpha}=-105.0\pm3.5$ milliarcsec year$^{-1}$ 
and $\mu_{\delta}=-1750\pm3.5$ milliarcsec year$^{-1}$,
computed as described in {Munn} {et~al.} (2004) by combining astrometry 
from the SDSS and from the USNO-B1.0 catalog ({Monet} {et~al.} 2003).
\rwd\ is present in seven Palomar Observatory Sky Survey (POSS) plates
and is present in one published SDSS observation.
A selection of these images is shown in Figure \ref{f_imgs}.
\rwd\ is catalogued in USNO-B1.0 as 1312-0217226, with proper motions of
$\mu_{\alpha}=-106\pm2$ milliarcsec year$^{-1}$ and
$\mu_{\delta}=-1744\pm1$ milliarcsec year$^{-1}$ 
based on 5 photographic epochs only.
We adopt a total proper motion of $\mu=1.75$ arcsec year$^{-1}$.
In Galactic coordinates, \rwd\ is located above the Galactic plane near 
anticenter ($l,b=174\arcdeg,63.5\arcdeg$).  Its proper motion is predominantly
in the $+l$ direction.  Taken together, those findings mean that its velocity
parallel to the Galactic plane is less than the Sun's (see \S \ref{pureH}).

Despite its presence in the USNO-B1.0 catalog,
this object has not been reported in the literature as a high-proper-motion
star.  No sources are listed in the SIMBAD or NED databases
within 1\,$\arcmin$ of its position.
\rwd\ is bright enough that it could have been found by
{Luyten} (1974).
   It was missed during examination of USNO-B
   high-proper-motion stars by {Levine} (2005) because
   it is 0.3 mag fainter than the limit of $R<18.0$ used in that study.
It should also have been found in the LSPM-N catalog ({L{\'e}pine}, {Shara}, \& {Rich} 2003; {L{\'e}pine} \& {Shara} 2005),
although it is near the lower magnitude limit and upper proper motion limit
of that catalog.
It may have been missed by previous searches because the image on the POSS-I E
plate appears double (Figure 1, top left panel).  This is likely due to a 
plate flaw, since the second component of the putative double 
is narrower than the point spread function.

Available photometry from published observations of \rwd\ is given in 
Table \ref{t_phot}.  The object is consistently faint in blue passbands.
There is a faint $J$-band feature in the Two Micron All Sky Survey (2MASS;
{Skrutskie} {et~al.} 2006) within 2$\arcsec$ of the expected position of \rwd\ at that
epoch.
To measure the flux in this feature,
we retrieved the 2MASS Atlas images covering this object,
measured 3$\arcsec$ radius aperture magnitudes in $J$, $H$ and $K_s$ at the
position of the potential $J$-band detection, and calculated appropriate
magnitude uncertainties.\footnote{The calculation of the magnitude uncertainties
followed the procdure in section VI.8.a.ii of the 2MASS Explanatory Supplement 
at http://www.ipac.caltech.edu/2mass/releases/allsky/doc/explsup.html.}  
These measurements are included in Table \ref{t_phot},
but their large uncertainties make them of limited use.
To obtain better near-infrared constraints, we obtained 
$JHK$ photometry of J1102+4113 on 2007 December 21 using the United
Kingdom Infrared Telescope (UKIRT) Fast-Track Imager (UFTI; {Roche} {et~al.} 2003) in
service mode.  The data were reduced in the standard fashion and the photometry
was calibrated using observations of UKIRT faint standard \#130 ({Leggett} {et~al.} 2006).
These measurements are also included in Table \ref{t_phot}.

\section{Spectra} \label{spec}

\begin{figure} \epsscale{1.20} \plotone{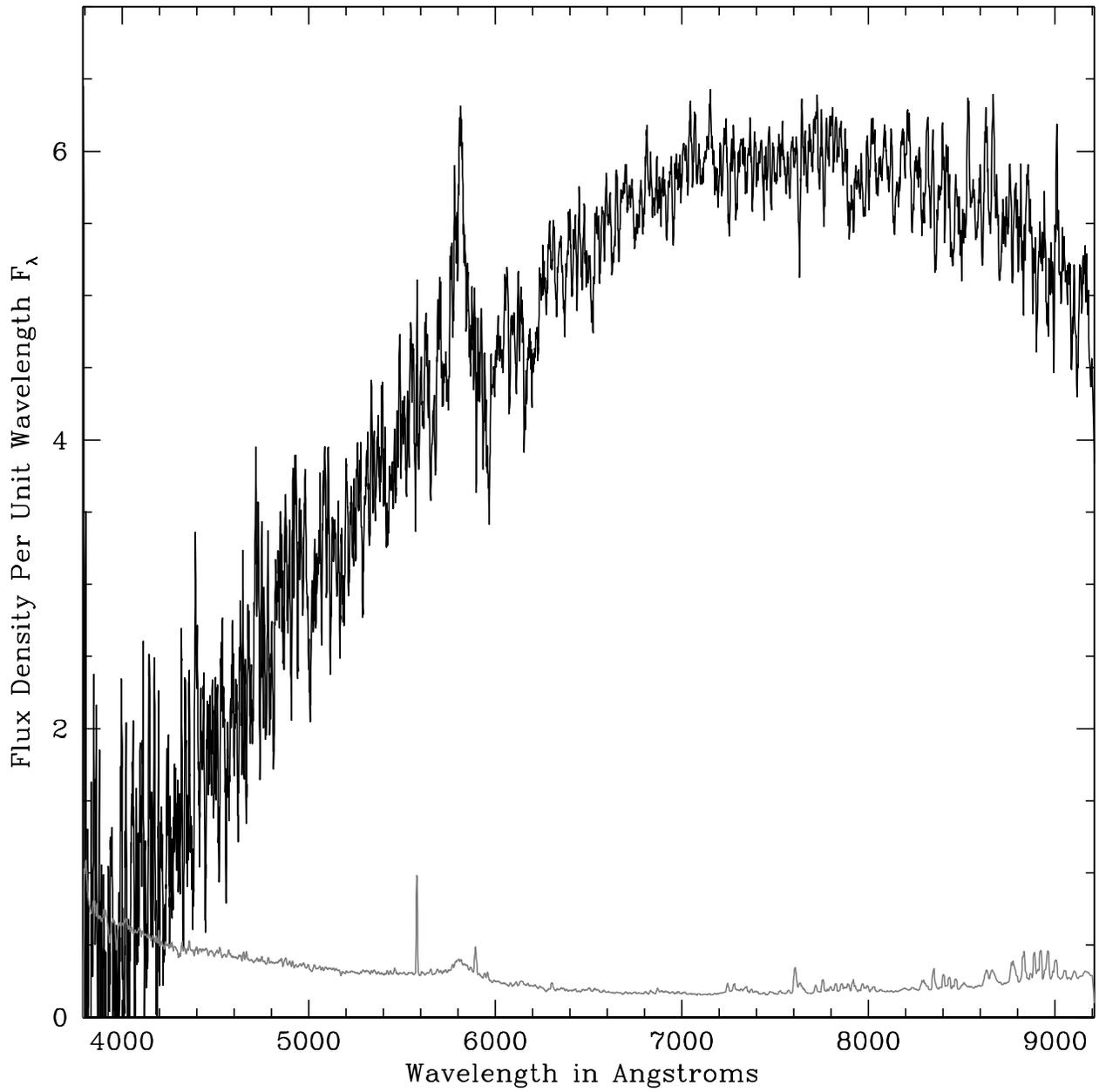} 
\caption{Full SDSS spectrum of \rwd, smoothed by a 7 pixel boxcar, plotted as
$F_\lambda$ (in units of 10$^{-17}$ erg s$^{-1}$ cm$^{-2}$ \AA$^{-1}$)
versus wavelength in \AA.
The uncertainty at each pixel is plotted along the bottom.
}\label{f_lam} \end{figure}

\begin{figure} \epsscale{1.20} \plotone{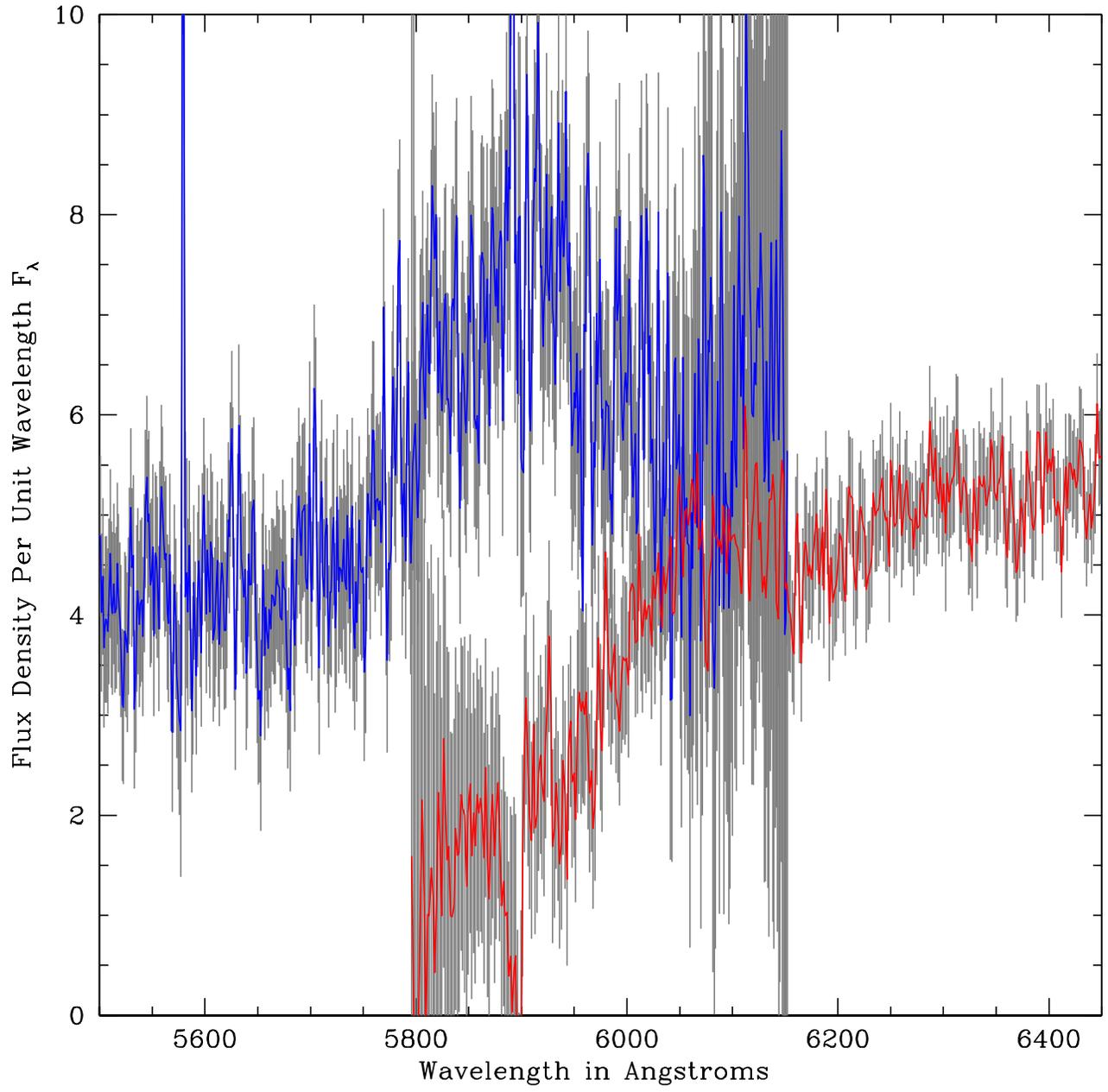} 
\caption{Average spectra of \rwd\ from the blue-wavelength SDSS spectrograph
(in blue) and the red-wavelength SDSS spectrograph (in red), along with
$\pm 1\sigma$ error bars (in gray).  The discrepant flux levels in the two
spectra between 5800--6000 \AA\ cast doubt on the reality of the emission
feature at 5750-5900 \AA\ in the combined spectrum.}\label{f_both} \end{figure}

Based on initial (`TARGET') photometric reductions of SDSS imaging,
\rwd\ was targeted as a high-redshift quasar candidate (target flag QSO\_HIZ).
A single SDSS spectrum was obtained on Modified Julian Date (MJD) 
53046 on SDSS spectroscopic plate No. 1437 and fiber No. 428.
In Figure \ref{f_lam} we present the full SDSS spectrum of \rwd.  Its spectrum
is red and featureless (in particular, there is no sign of Balmer absorption)
except for a possible broad emission feature at 5750--5900 \AA.  This feature
is near one end of the wavelength region 5800--6150 \AA\ where both the blue
and red SDSS spectrographs record flux from an object.
To investigate the reality of this feature, we examined the four individual
blue and red exposures which were all combined to
produce the final weighted average SDSS spectrum shown in Figure \ref{f_lam}.
The broad feature is present
in each of the four individual blue exposures,
but not in any of the four individual red exposures.
In fact, the red exposures all show a dip in flux below 6050 \AA.
In Figure \ref{f_both} we present the average spectra of this object obtained
with the blue and red SDSS spectrographs, shown in blue and red respectively,
along with the $\pm 1\sigma$ uncertainties for both spectra, shown in gray.
The flux levels at 5800--6000 \AA\ in the two spectra do not agree
within the uncertainties.  This disagreement is indicative of some problem
with the SDSS spectrum in this wavelength range.
We conclude that the broad feature at 5750-5900 \AA\ is an 
artifact,\footnote{Some pixels at nearby wavelengths are flagged as potentially
being untrustworthy, with some or all of the flags {\sc NEARBADPIXEL, LOWFLAT,
SCATTEREDLIGHT, BADFLUXFACTOR}, and {\sc BADSKYCHI} ({Stoughton} {et~al.} 2002).  
Empirically, we have found that the grow radius around such untrustworthy
pixels is, on rare occasions, not large enough to flag all apparently
problematic pixels in SDSS spectra.  This spectrum may be one such case.}
although an independent spectrum would still be worth obtaining to verify that
conclusion.

There is no other statistically significant absorption or emission feature in
the spectrum.  In particular, 
there is no sign of H$\alpha$ absorption to a $3\sigma$ limit of $\sim$0.8~\AA.
(The smoothed spectrum shows a dip in flux at 6522~\AA, but it is due to two
noise spikes which are narrower than the instrumental resolution.)

\section{Analysis and Comparison with Models} \label{think}

We classify \rwd\ as a DC WD, since its optical spectrum has no robust features.
   The featureless spectrum of J1102+4113 is similar to other cool DC WDs.
   At optical wavelengths, the reddest of these are
   WD~0346+246 ({Oppenheimer} {et~al.} 2001b), GD~392B ({Farihi} 2004), WD~1247+550 ({Liebert}, {Dahn}, \& {Monet} 1988)
   and WD~1310-472 ({Bergeron}, {Leggett}, \& {Ruiz} 2001).  These four stars all have $B-V \simeq 1.4$
   and $1.39 < V-I < 1.46$.\footnote{Two WDs reported to have $V-I>1.5$ by
{Oswalt} {et~al.} (1996) in fact have $V-I < 1.0$ ({Bergeron} {et~al.} 2001; Harris, unpublished).}
Only one of these four WDs has a measured $g-i$ color, but the small dispersion
in their $V-I$ colors means there will be a small dispersion in their $g-i$ 
colors.
Thus, comparing the $g-i = 1.67$ measured for WD~1247+550
with the $g-i = 1.99$ measured for \rwd, we conclude that \rwd\ is $\simeq 0.3$
	mag redder in $g-i$ than the reddest previously known DC white dwarfs.
The only other non-magnetic WDs\footnote{No known magnetic WD has a red,
featureless spectrum like that of \rwd.  Nonetheless, polarimetric observations
would be useful to ensure that \rwd\ is not some sort of unusual magnetic WD.}
known to have optical colors approaching those
of \rwd\ are DZ WDs with extremely strong calcium and sodium absorption:
WD~2251-070 ({Liebert} {et~al.} 1988),
WD~J2356$-$209 ({Oppenheimer} {et~al.} 2001a) and SDSS J133001.13+643523.8 ({Harris} {et~al.} 2003).
In contrast, \rwd\ shows no evidence of absorption from
\ion{Na}{1} $\lambda\lambda$5891,5897, \ion{Ca}{1} $\lambda$4227\,\AA\ or
\ion{Ca}{2} $\lambda\lambda\lambda$8500,8544,8664\,\AA.
A cool atmosphere with an extremely low metal abundance
seems the best explanation for its red color.

For a surface temperature as low as that indicated by the red color of \rwd,
both pure hydrogen and pure helium model atmospheres
are expected to be essentially featureless in the optical.
In cool, pure H atmospheres, most H is in the form of H$_2$ and most
atomic H is in the ground state, leading to negligible Balmer absorption
(which would be extremely pressure broadened in any case).
In cool, pure He atmospheres, most He is neutral and the lower levels of
optical transitions of \ion{He}{1} are not populated for 
\teff $\lesssim 12,000$~K, leading to highly optically transparent atmospheres.
Nonetheless, there can be detectable differences between the spectra of
hydrogen- and helium-dominated cool WDs, particularly in the near-infrared
where collision-induced absorption is affected by the composition.

To constrain the atmospheric composition of \rwd, we fit model WD atmospheres 
to the UKIRT photometry and to the SDSS photometry converted to an AB system.
The SDSS magnitudes are already on an AB system to within $\simeq 1$\% in 
$g$ and $r$, while in other bands we applied the corrections
$u_{AB} = u_{SDSS} - 0.04$, $i_{AB} = i_{SDSS} + 0.02$ and
$z_{AB} = z_{SDSS} + 0.03$
({Abazajian} {et~al.} 2004; {Eisenstein} {et~al.} 2006; cf. {Holberg} 2007; 
Holtzman et al. 2008, in preparation).  To account for the $\simeq 1$\% 
uncertainties in these corrections, we increased the uncertainties from 2\% to
3\% for each SDSS magnitude besides $u$ in the fitting.

\begin{figure}\epsscale{0.58} \includegraphics[scale=0.89,angle=0]{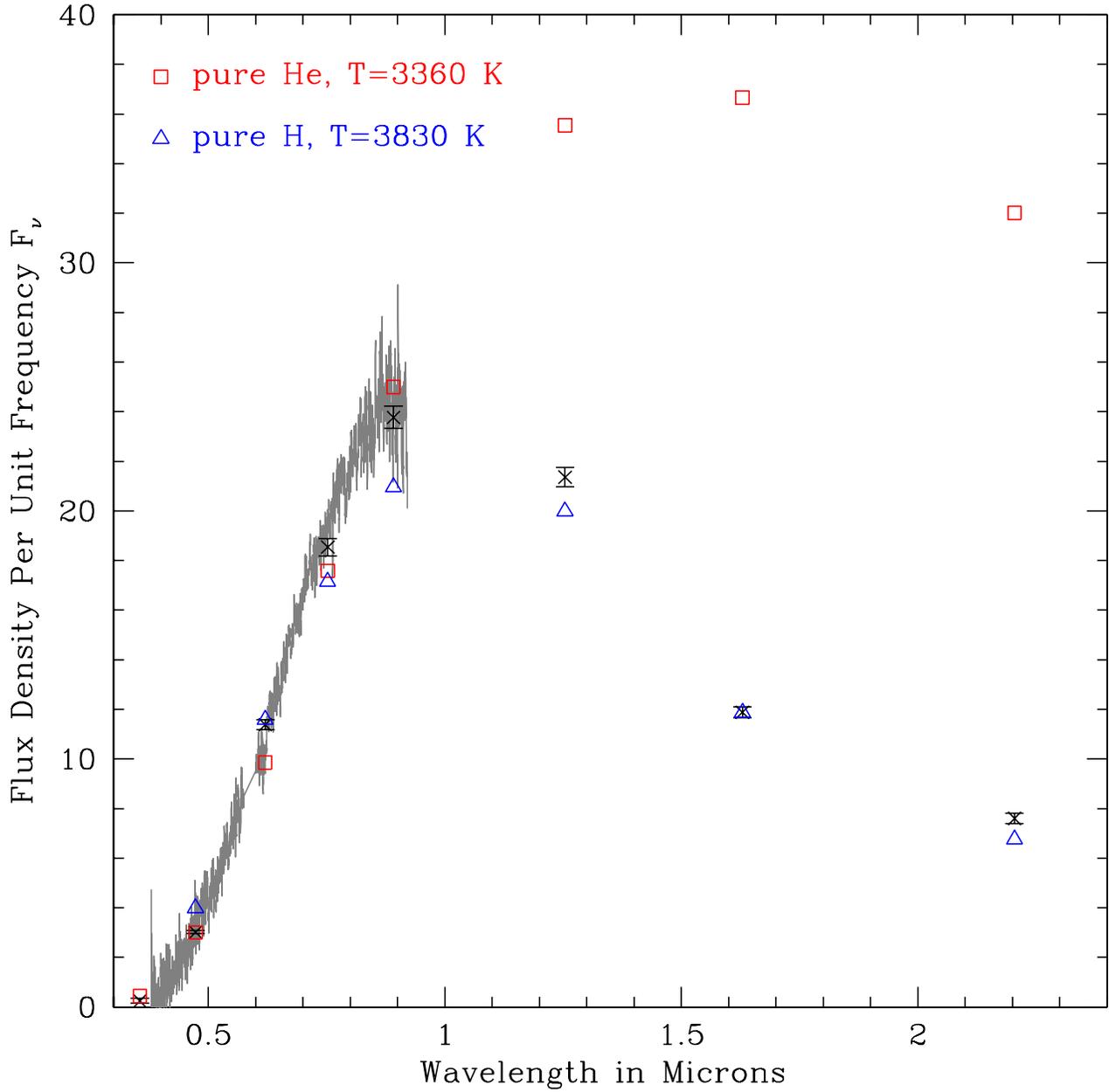}
\caption{Observed spectrum (plotted as $F_\nu$ in units of erg s$^{-1}$
cm$^{-2}$ Hz$^{-1}$ versus wavelength in microns),
observed photometry and synthetic model photometry of \rwd.
The SDSS spectrum is shown as the dark gray line and the associated
uncertainties as the light gray line.  The emission feature we believe to be 
spurious has been interpolated over.  
The observed photometry of \rwd\ is shown as the black error bars.
Blue triangles are synthetic photometry from the best-fit pure-H model.
Red squares are synthetic photometry from a pure-He model fit to the optical
data only.}\label{rwdfnu3} \end{figure}

In Figure \ref{rwdfnu3} we plot the SDSS spectrum of \rwd, the SDSS and UKIRT 
photometry and synthetic photometry from fits discussed below,
all in units of flux density per unit frequency versus wavelength.  
For illustrative purposes, the spectrum has been scaled upwards by 0\fm6
to match the optical photometry.  Such scaling is needed because not all
the light from the object is captured by the spectroscopic fiber.  
The scaling is greater than the typical SDSS value of
0\fm35 ({Adelman-McCarthy} {et~al.} 2008), probably because the fiber placement did not account for the
1\farcs4 proper motion of \rwd\ between the imaging and spectroscopic epochs.

\subsection{Comparison with Pure Hydrogen Models} \label{pureH}

We first fit pure-hydrogen atmosphere models ({Kowalski} 2006b; {Kowalski} \& {Saumon} 2006; {Kowalski} 2007) to the
photometry of \rwd\ (excluding the $u$ band,
whose uncertainties are too large to be useful),
assuming a typical WD gravity of $\log g = 8$ (e.g., {Fontaine}, {Brassard}, \&  {Bergeron} 2001).
The resulting fit yields $\tiff=3830$~K.  Synthetic photometry from this model
is shown as the blue triangles in Figure \ref{rwdfnu3}.
The fit is reasonable given the model uncertainties, although the formal
$\chi^2$ is quite poor: $\chi^2=111$ for 4 degrees of freedom
($\nu=n-1=6$ minus \teff\ minus a normalization factor), with
about $\sim$75\% of the $\chi^2$ signal coming from the discrepancy at $g$.
The red optical colors are produced in the far red wing of Ly$\alpha$
(perturbed at high density), while the dips in the near-IR spectrum are
due to H$_2$$-$H$_2$ collision-induced absorption (CIA).

While \rwd\ is 0\fm22 redder in $g-z$ than any pure-H model, it is an open
question how well such models reproduce the relevant physics at 
$\tiff\simeq 3800$~K.  
Hydrogen in the atmosphere of a $\tiff\simeq 3800$~K WD is approaching a
density where nearby molecules are strongly correlated and where the
refractive index is significantly greater than unity.  Furthermore, there 
are still uncertainties about the reliability of H$_2$ CIA opacities in WDs.
Observationally, there are few WDs known near that temperature
with which comparisons with models can be made.
At $\tiff\lesssim 3800$~K, CIA will affect the optical colors of WDs and their
$g-z$ colors will become bluer.  However, given all the physical effects
discussed above that may not be handled correctly in the models, it is not
certain how optically red pure-H atmospheres become before they turn bluer.
In short, the models could be sufficiently in error to explain the formally
poor fit of the pure H model predictions to the \rwd\ data.

The distance to the WD in our pure-H model fit is 35 pc,
making the tangential velocity $v_{tan}=290$ km s$^{-1}$.  The cooling time
to reach \teff=3830 K is $\approx 9.6$ Gyr, assuming a typical WD mass
of 0.6 $M_\odot$ with a core of equal parts C and O ({Fontaine} {et~al.} 2001).  Including
$\approx 1$ Gyr for its progenitor's lifetime ({Fontaine} {et~al.} 2001), for a pure-H model
\rwd\ is the remnant of a star that formed $\approx 11$ Gyr ago.
Alternatively, at the same temperature, 
if $\log g = 9$ (9.5),
then \rwd\ would have the following approximate properties:
$M=1.11$ (1.36) $M_\odot$, $d=17$ (10) pc, $v_{tan} = 140$ (83) km s$^{-1}$,
cooling time 7 (3) Gyr and progenitor lifetime $\ll 1$ Gyr.
If \rwd\ has $\log g = 7.5$, then it would be at $d=42$~pc with 
$M=0.35 M_\odot$ and its total age would
be greater than the age of the universe unless it is an unresolved double
degenerate or a product of common-envelope binary star evolution ({Fontaine} {et~al.} 2001).

We can calculate the components of the object's space velocity relative
to the Sun, $U$, $V$ and $W$, which are positive in the directions of
Galactic Center, Galactic rotation, and the North Galactic Pole, respectively.
If we assume zero radial velocity we find ($U,V,W$) = 
(63,$-$280,46) km s$^{-1}$ for $d=35$ pc,
(30,$-$140,22) km s$^{-1}$ for $d=17$ pc,
and (18,$-$80,13) km s$^{-1}$ for $d=10$ pc.
{Chiba} \& {Beers} (2000) give ($<$$U$$>$,$<$$V$$>$,$<$$W$$>$) =
(17$\pm$141, $-$187$\pm$106, $-$5$\pm$94) km s$^{-1}$ for the halo
and (4$\pm$46, $-$20$\pm$50, $-$3$\pm$35) km s$^{-1}$ for the thick disk.
Thus, the space velocities of \rwd\ are most consistent with those of the halo.
However, if \rwd\ is at $d\lesssim 13.5$ pc, it could be a young, thick disk
WD with an unusually (but not unprecedentedly) high mass.
A parallax measurement is needed to discriminate between these possibilities
and to pin down the mass of \rwd.

\subsection{Comparison with Pure Helium Models} \label{pureHe}

\begin{figure}\epsscale{0.42} \includegraphics[scale=0.65,angle=-90]{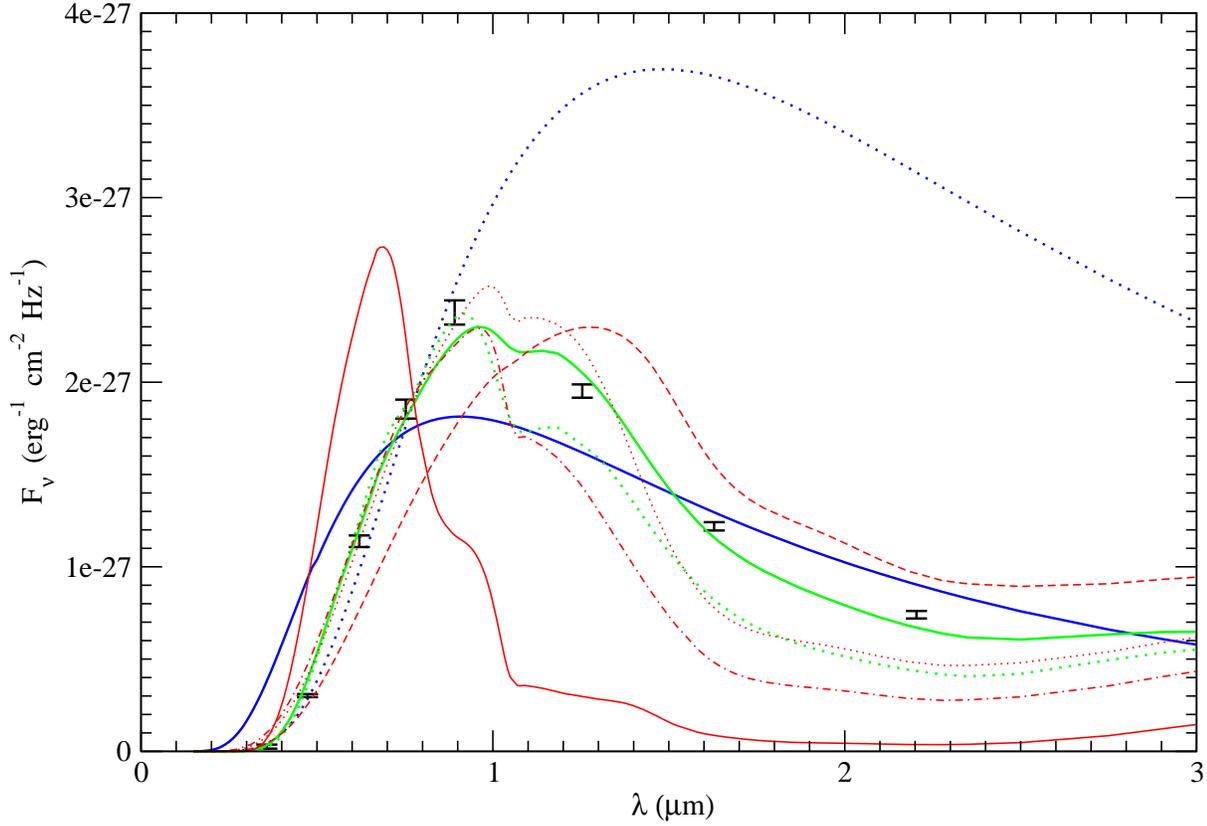}
\caption{Synthetic spectra of pure H, pure He and mixed H/He models, 
all assuming $\log g=8$.  
Fits to all data yield \teff=3830 K for pure H (solid green)
and \teff=5381 K for pure He (solid blue).
Fits to optical data only yield \teff=3450 K for pure H (dotted green)
and \teff=3360 K for pure He (dotted blue).
Helium-dominated models with \teff=3500 K are shown by the red lines,
with hydrogen contents as follows: log H/He = $-$3 (solid), $-$4 (dash-dotted),
$-$5 (dotted) and $-$5.5 (dashed).}\label{wd_mixed} \end{figure}

We next fit pure He models ({Kowalski} \& {Saumon} 2004, 2006) to the data (again assuming
$\log g = 8$), but were not able to find a realistic fit.
If the near-infrared photometry is omitted, a good fit is found with
\teff=3360 K.  Synthetic photometry from this model is shown as the red 
squares in Figure \ref{rwdfnu3}; the model near-infrared fluxes greatly
overpredict the observed fluxes.  Thus, a good fit cannot be found to the
combined optical and near-IR photometry of \rwd\ with a pure He model:
if the temperature is increased to move the peak of the fit to
shorter wavelengths, the slope of the optical fit will deviate unacceptably
from the observations (see the blue lines in Figure \ref{wd_mixed}).

\clearpage

\subsection{Comparison with Mixed Hydrogen-Helium Models} \label{mixedHHe}

We also consider mixed H/He models, as even a tiny amount of hydrogen
in a helium-dominated atmosphere
(${\rm H/He}\gtrsim 10^{-10}$) can induce sufficient H$_2$$-$He CIA to cause a
significant near-infrared flux deficit (see, e.g., Figure 5 of {Bergeron} \& {Leggett} 2002).
The results of normalizing mixed H/He models with $\tiff=3500$~K to the
photometry are shown in Figure \ref{wd_mixed}; for slightly higher or lower 
\teff\ ($\pm$500~K) the overall results are similar.  None of the mixed H/He
models yield a better match to the data than the pure hydrogen model, although
a model with ${\rm H/He}=10^{-5}$ fits at wavelengths $\lambda<1.5$ microns.

In summary, the poor fit of pure-He models to the photometry means that 
there is certainly hydrogen in the atmosphere of this WD.
However, the H/He ratio cannot be pinned down with current models,
due to uncertainties in how accurately the models treat the complicated physics
involved (refraction, non-ideal equation of state, non-ideal chemistry,
perturbed Lyman series absorption, etc.; see {Kowalski} 2006a).
Both H/He$\gg$1 and H/He$\ll$1 are possible.

\subsection{Modeled and Observed Fluxes and Colors} \label{compare}

If only observational uncertainties are considered, 
no current WD model atmosphere provides a formally acceptable fit
to the photometry of \rwd.  To guide future models,
it is worth examining at what wavelengths the discrepancies arise.

In both Figures \ref{rwdfnu3} and \ref{wd_mixed}, it can be seen that
\rwd\ is redder at optical wavelengths than most model fits.  To even 
approximately match both the red optical colors and the depression 
of the near-infrared flux (relative to a blackbody of the same temperature)
requires some hydrogen in its atmosphere.

In terms of $u-g$ vs. $g-z$,
\rwd\ matches neither pure-H models nor
pure-He models (see Table~\ref{t_phot} and Figure 4b of {Kowalski} \& {Saumon} 2006).
It is redder in $g-z$ than pure-H models at any $u-g$
but is bluer in $u-g$ than the pure-He models at its $g-z$ color.
Because of the large uncertainty in its $u-g$ color, \rwd\ is only $1.7\sigma$
redder in $u-g$ than the pure-He sequence of {Kowalski} \& {Saumon} (2006).

\section{Summary} \label{end}

SDSS J110217.48+411315.4 is a cool WD with either a pure hydrogen atmosphere at
$\tiff\simeq 3830$~K or a mixed hydrogen-helium atmosphere with
${\rm H/He}\simeq 10^{-5}$ and $\tiff\simeq 3500$~K.  In either case,
its distance is expected to be $d\lesssim 40$~pc.  At its best-fit distance of
$d\simeq 35$~pc, its high proper motion of 1.75$\arcsec$/year makes it a member
of the halo, with an estimated total age of $\approx 11$~Gyr.  If its distance
is $d\lesssim 13.5$ pc, it could be a halo or a thick disk object.  A parallax
measurement for \rwd\ is needed to settle the question of its origin and to
determine its mass.  To improve the data available for fitting models to \rwd,
deeper $u$ photometry and photometry at $\lambda>2.5$ microns might be useful,
as might optical spectroscopy with a higher signal-to-noise ratio.  As for the
models themselves, improvements to mixed H$-$He models are known to be needed.
In addition, better H$_2$ CIA opacities (especiallly at optical wavelengths)
and more examples of WDs at $\tiff\approx 4000$~K are needed to determine how
well, on average, pure-H models at such temperatures match real WD atmospheres.

In principle, the SDSS can be used to place an upper limit on the surface
density of red, high-proper-motion WDs like J1102+4113.
In practice, additional spectroscopy beyond that obtained by the SDSS
will be needed to place such a limit, because objects like J1102+4113 have
colors too similar to those of the stellar locus to be routinely selected
for SDSS spectroscopy.  (J1102+4113 itself barely qualified as a quasar 
candidate based on preliminary (`TARGET') SDSS imaging reductions, and no 
longer qualified for spectroscopy based on final (`BEST') reductions.)
J1102+4113 is the only $g<19.7$, $g-r>1.4$ object in the 8417 deg$^2$ 
SDSS DR6 Legacy imaging database with SDSS+USNO-B1.0 proper motion
$\mu\geq 1.5\arcsec$/year ({Munn} {et~al.} 2004),
but similarly red objects with $\mu\lesssim 1\arcsec$/year that lack SDSS
spectroscopy exist in the database.  Whether or not any of those objects
turn out to be WDs, J1102+4113 will remain one of the highest proper motion
cool WDs in the sky as seen from Earth.

\acknowledgments
We thank D. Saumon and the anonymous referee for helpful comments.
PBH was supported by NSERC and AA by the York-Seneca Summer Science and 
Technology Program.  PMK acknowledges partial support from Ruhr Universit\"at. 
SKL's research is supported by the Gemini Observatory, which is operated by the
Association of Universities for Research in Astronomy, Inc., on behalf of the
international Gemini partnership of Argentina, Australia, Brazil, Canada, 
Chile, the United Kingdom, and the United States of America.

Some of the data reported here were obtained as part of Service Programme 
1771 at the United Kingdom Infrared Telescope, which is operated by the Joint
Astronomy Centre on behalf of the Science and Technology Facilities Council of
the U.K.  This research has also made use of:
the NASA/IPAC Infrared Science Archive and 
the NASA/IPAC Extragalactic Database (NED), which are operated by the Jet Propulsion Laboratory, California Institute of Technology, under contract with the National Aeronautics and Space Administration;
the SIMBAD database, operated at CDS, Strasbourg, France;
data products from the Two Micron All Sky Survey, which is a joint project of the University of Massachusetts and the Infrared Processing and Analysis Center/California Institute of Technology, funded by the National Aeronautics and Space Administration and the National Science Foundation;
and the USNOFS Image and Catalogue Archive
operated by the United States Naval Observatory, Flagstaff Station
(http://www.nofs.navy.mil/data/fchpix/).

Funding for the SDSS and SDSS-II has been provided by the Alfred P. Sloan Foundation, the Participating Institutions, the National Science Foundation, the U.S. Department of Energy, the National Aeronautics and Space Administration, the Japanese Monbukagakusho, and the Max Planck Society, and the Higher Education Funding Council for England. The SDSS Web site is http://www.sdss.org/.
The SDSS is managed by the Astrophysical Research Consortium for the Participating Institutions. The Participating Institutions are the American Museum of Natural History, Astrophysical Institute Potsdam, University of Basel, University of Cambridge, Case Western Reserve University, The University of Chicago, Drexel University, Fermilab, the Institute for Advanced Study, the Japan Participation Group, The Johns Hopkins University, the Joint Institute for Nuclear Astrophysics, the Kavli Institute for Particle Astrophysics and Cosmology, the Korean Scientist Group, the Chinese Academy of Sciences, Los Alamos National Laboratory, the Max-Planck-Institute for Astronomy, the Max-Planck-Institute for Astrophysics, New Mexico State University, Ohio State University, University of Pittsburgh, University of Portsmouth, Princeton University, the United States Naval Observatory, and the University of Washington.

\clearpage

\setlength{\tabcolsep}{0.02in}
\tabletypesize{\scriptsize}
\begin{deluxetable}{lrcccccccccc}
\tablecaption{Photometry of SDSS J110217.48+411315.4\label{t_phot}}
\tabletypesize{\scriptsize}
\tablewidth{505.00000pt}
\tablehead{
\colhead{Source} & \colhead{MJD} &
\colhead{$u\pm\sigma_u$} & \colhead{$g\pm\sigma_g$} & \colhead{$r\pm\sigma_r$} &
\colhead{$i\pm\sigma_i$} & \colhead{$z\pm\sigma_z$} &
\colhead{$J\pm\sigma_J$} & \colhead{$H\pm\sigma_H$} & 
\colhead{$K\pm\sigma_{K}$} & \colhead{$u-g$} & \colhead{$g-z$} 
}
\startdata	
POSS1& 35183.7938& \nodata & $20.59$ & \nodata & \nodata & \nodata & \nodata & \nodata & \nodata & \nodata & \nodata \\
POSS1& 35183.8321& \nodata & \nodata & $18.58$ & \nodata & \nodata & \nodata & \nodata & \nodata & \nodata & \nodata \\
POSS2& 47540.9117& \nodata & \nodata & $18.67$ & \nodata & \nodata & \nodata & \nodata & \nodata & \nodata & \nodata \\
POSS2& 48001.6992& \nodata & $20.26$ & \nodata & \nodata & \nodata & \nodata & \nodata & \nodata & \nodata & \nodata \\
POSS2& 48294.8804& \nodata & unavailable & \nodata & \nodata & \nodata & \nodata & \nodata & \nodata & \nodata & \nodata \\
POSS2& 50094.9138& \nodata & \nodata & \nodata & $18.41$ & \nodata & \nodata & \nodata & \nodata & \nodata & \nodata \\
2MASS& 50912.8346& \nodata & \nodata & \nodata & \nodata & \nodata & 17.78$\pm$0.76 & 17.37$\pm$0.92 & 17.52$\pm$1.81 & \nodata & \nodata \\ 
POSS2& 50948.6813& \nodata & \nodata & unavailable & \nodata & \nodata & \nodata & \nodata & \nodata & \nodata & \nodata \\
 SDSS& 52754.1402& 23.01$\pm$0.48& 20.20$\pm$0.02& 18.76$\pm$0.02& 18.21$\pm$0.02& 17.93$\pm$0.02& \nodata & \nodata & \nodata & 2.81$\pm$0.48 & 2.27$\pm$0.03\\
UKIRT& 54456.1192& \nodata & \nodata & \nodata & \nodata & \nodata & 17.24$\pm$0.02 & 17.33$\pm$0.02 & 17.34$\pm$0.03 & \nodata & \nodata \\ 
\enddata
\tablecomments{Optical magnitudes are on the AB system (but see text),
and 2MASS and UKIRT magnitudes on the Vega system.
2MASS magnitudes use the 2MASS filter system (Skrutskie et al. 1997)
and UKIRT magnitudes use the MKO filter system 
({Ghinassi} {et~al.} 2002).
POSS magnitudes have been recalibrated to the indicated SDSS
filter ({Sesar} {et~al.} 2006).  
SDSS magnitudes are point-spread-function (PSF) 
magnitudes, but have been converted from asinh magnitudes ({Lupton}, {Gunn}, \& {Szalay} 1999)
to traditional magnitudes,  
using the information in Table 21 of {Stoughton} {et~al.} (2002).
For this object, the difference is significant only in the $u$ band.}
\end{deluxetable}

\end{document}